\documentstyle[floats,aps,prl,epsfig]{revtex}

\begin{document}
\draft
\twocolumn[\hsize\textwidth\columnwidth\hsize\csname@twocolumnfalse%
\endcsname
\title{Excitonic emission in the presence of a two dimensional electron gas:\\ a
microscopic understanding}
\author{Y. Yayon, M. Rappaport, V. Umansky and I. Bar-Joseph}
\address{Department of Condensed Matter Physics, The Weizmann Institute of Science,
Rehovot 76100, Israel}
\maketitle

\begin{abstract}
The near- and far-field photoluminescence spectra of a gated two-dimensional
electron gas have been measured. Spatial fluctuations in the electron
density are found to be manifested as spatial fluctuations in the emission
amplitude of the negatively charged exciton (X$^{-}$) and in the peak energy
of the neutral exciton (X). Consequently, the far-field X$^{-}$ spectrum is
a homogoneously broadened Lorentzian, while the X lineshape exhibits
substantial inhomogenous broadening. We present a novel and simple technique
to extract the electron density from the X$^{-}$ spectrum. We show that
width of the far-field X line is proportional to the electron density
fluctuations; hence it can be used to characterize the inhomogeneity in the
electron density.
\end{abstract}
]

The photoluminescence (PL) spectrum of a low density two
dimensional electron gas (2DEG) has been the subject of
considerable interest in recent years. It is observed that at a
certain critical density, which depends on sample parameters but
is typically between $10^{10}-10^{11}$ cm$^{-2}$, the spectrum
abruptly changes, from a broad line at high densities to two
narrow peaks at low densities. A number of spectroscopy
experiments at zero and high magnetic fields have clearly shown
that the two peaks are associated with the neutral (X) and
negatively charged (X$^{-}$) excitons \cite
{Kheng,GlebPRL,Shields}. The X$^{-}$ is formed by the binding of a
photoexcited electron-hole pair to an electron of the 2DEG. When
the electron density is decreased its relative intensity
decreases, and the spectrum becomes dominated by the X line. It
was found that the appearance of this excitonic spectrum is
correlated with a rapid decrease in the 2DEG conductivity
\cite{GlebPRL}. This drop in conductivity was theoretically
predicted to mark the onset of strong localization: As the 2DEG
density is decreased, it becomes ineffective in screening the
potential fluctuations due to the remote ionized donors, and they
grow substantially \cite {Efros,Nixon}. Consequently, the
electrons become localized in those potential fluctuations, and
their ability to screen is further reduced. It was suggested that
this reduction in screening and the subsequent strong
localization of the electrons, allowed the observation of
excitons in the presence of the 2DEG \cite{GlebPRL}. This
conclusion was corroborated by a near-field optical study, which
has shown that the X$^{-}$ intensity exhibits strong spatial
fluctuations: regions with high (low) electron density give rise
to a strong (weak) X$^{-}$\ signal \cite{EytanPRL}.

Our goal in the present work is to establish a quantitative{\it \ }relation
between the optical spectrum at low electron density and the properties of
the 2DEG. We measure and carefully analyze both the near- and far-field PL
spectra of a gated 2DEG. We find that the near-field X$^{-}$ lineshape is a
Lorentzian, with an amplitude that varies strongly between different points
in the sample, but with a very narrow distribution of peak energies and
widths. Consequently, the far-field X$^{-}$ lineshape is a homogeneously
broadened Lorentzian, with a numerator that is proportional to the average
density. We show that one can easily determine the average electron density
in this low density regime from the PL spectrum. The near-field spectra of
the X, on the other hand, exhibit a broad distribution of peak energies. We
show that there is a clear correlation between the 2DEG density fluctuations
and those of the X peak energies: regions with high (low) electron density
give rise to a high (low) X peak energy. Hence, we conclude that the
electron density fluctuations are the origin of the inhomogeneous broadening
of the far-field X peak. Thus, the far-field X width is a measure for the%
{\it \ }electron density fluctuations{\it \ }in the sample.

The sample we studied is a 20 nm GaAs quantum well (QW). A $3.6$ nm Si doped
donor layer with a density of $4\times 10^{18}$ cm$^{-3}$ is separated from
the QW by a 50 nm Al$_{\text{0.36}}$Ga$_{\text{0.64}}$As spacer layer. The
distance between the QW and the sample surface is $107$ nm. A $2\times 2$ mm$%
^{2}$ mesa was etched, and ohmic contacts were alloyed to the 2DEG layer. A
4.5 nm PdAu semitransparent gate was evaporated on top of the sample. The
ungated 2DEG density and mobility (after illumination) at $4.2$ K are $%
2.2\times 10^{11}$ cm$^{-2}$, and $3\times 10^{6}$ cm$^{2}/$V $\sec $,
respectively.

Near-field PL measurements were performed using a system, which is described
in detail in Ref. \cite{NSOM Ultramicroscopy}. It is a low temperature
near-field scanning optical microscope (NSOM) that operates in the
collection mode: the sample is illuminated uniformly by a single-mode fiber
and the emitted PL is collected through a tapered Al-coated optical fiber
tip. The tip clear aperture diameter and transmission were measured to be $%
200$ nm and $10^{-3}$, respectively. The spatial resolution was determined
using a grating mask to be $180$ nm. The tip is glued to a quartz tuning
fork, and its piezoelectric signal is used to control the height of the tip
above the sample surface at $\sim 10$ nm \cite{Karrai}. For the excitation
we used a He:Ne laser at 632.8 nm. The collected PL was dispersed by a 0.5 m
spectrometer and detected using a cooled CCD camera. The overall system
spectral resolution is 0.038 nm.

\begin{figure}[t]
\centering\epsfig{file=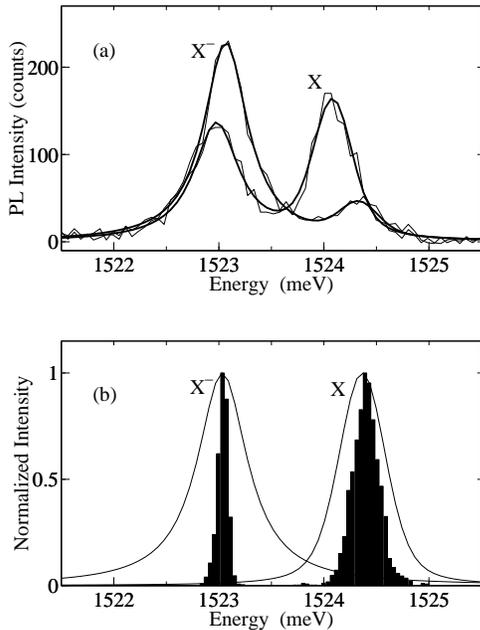,height=8.5cm,angle=0}\caption{a)
Two near-field PL spectra (thin lines) and fitted curves (thick
lines) at two points, 0.25 $\mu $m away from each other. b)
Histograms of the peak energy of the X$^{-}$ and X lines together
with the fitted far-field lineshape for each excitonic line. The
measurement is at a gate voltage that corresponds to
$n_{\text{e}}=3.1\times 10^{10}$ cm$^{-2}$.}\label{fig1}
\end{figure}%
Figure \ref{fig1}a shows two near-field PL spectra (thin lines)
measured at a fixed gate voltage at two points in the sample,
$0.25$ $\mu $m\ apart. Both spectra consist of two peaks, X and
X$^{-}$. It is seen that there are large differences between the
two spectra over this small distance. The differences are
manifested in the relative heights, widths and peak energies of
each spectral line. The strong spatial fluctuations in the
X$^{-}$ intensity were already reported by us in a previous work
\cite{EytanPRL}. We have significantly improved the signal to
noise in the present measurements, and were thus able to better
resolve the properties of the emission spectra. In extracting
these properties we performed line fits at more than $3000$
points in a scanned area of $4\times 5.5$ $\mu $m$^{2}$, and
repeated this procedure at three gate voltages. The thick lines
in Fig. \ref{fig1}a are the fitted curves for the two spectra. The
lineshape we used to fit each peak is a Voigt function, a
convolution of a Lorentzian and a Gaussian, which describe
homogeneous and inhomogeneous broadening, respectively. Such a
lineshape was already used to fit the PL spectrum of an intrinsic
sample, in which a 2DEG was created by optical excitation
\cite{Menassen}. This is obviously a simplified lineshape, which
neglects the contribution of high order effects. Nevertheless, we
find that it gives a very good fit to the measured spectra. We
obtained a significantly better fit at all points (nearly 10,000
in all measurements) using a Lorentzian-dominated Voigt lineshape
for the X$^{-}$, with a negligible Gaussian width. The X line, on
the other hand, is best described by a Voigt function with a
significant inhomogeneous contribution.

Let us examine the resulting properties of the X$^{-}$ and X
lineshapes. Figure \ref{fig1}b shows the normalized histograms of
the peak energies at the $\sim 3000$ points for each excitonic
line together with the normalized fitted far-field lineshapes. It
can be clearly seen that the X$^{-}$ peak energy is nearly
independent of position with significantly less scatter than the
far-field linewidth. We also find a very narrow distribution of
the near-field X$^{-}$ linewidth (not shown), with an average
which is equal to the far-field width. We therefore conclude that
the far-field X$^{-}$ line is predominantly {\it homogeneously
}broadened{\it : }It is a sum of many local Lorentzians, each
having nearly the same peak energy and width but a different
amplitude. This summation results in a far-field lineshape that is
also a Lorentzian. The near-field X peak histogram, on the other
hand, exhibits significantly greater scatter in its peak
energies. This scatter is larger than the X homogenous linewidth
(Lorentzian width) and gives rise to {\it inhomogeneous
}broadening{\it \ }of the X far-field spectrum. Indeed, we
observe in Fig. \ref{fig1}b that the far-field X width is
comparable to the X peak energy distribution. We find that the X
peak is inhomogeneously broadened already in the near-field
measurements. This implies that the broadening mechanism occurs
on a length scale smaller than our spatial resolution, and the
width of the X peak energy histogram measured at higher spatial
resolution would be even larger. We shall address this point
later when we discuss the origin of the inhomogeneous broadening.

To check this assignment of broadening types we have conducted a set of
far-field PL measurements at various gate voltages and illumination
intensities (six laser intensities and 50 gate voltages per laser
intensity), and fitted the measured spectra with the double Voigt function
discussed above. We have found that throughout the gate voltage range the X$%
^{-}$ is well fitted by a Lorentzian (convolved with a Gaussian, which is
much narrower than the Lorentzian) while the X lineshape has substantial
inhomogeneous Gaussian broadening. These findings are consistent with the
results of the near-field measurements.

\begin{figure}[t]
\centering\epsfig{file=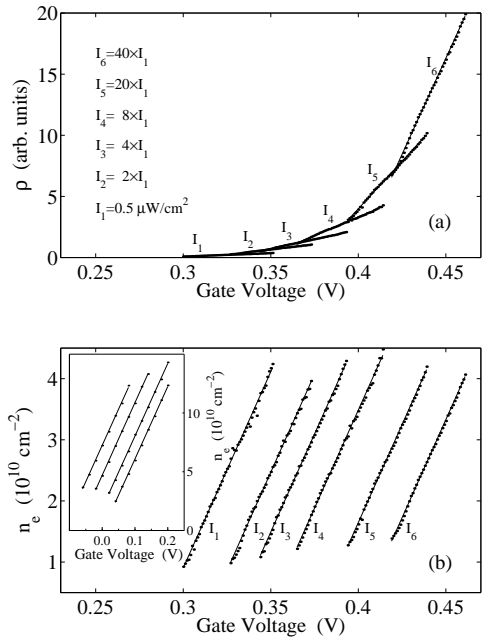,height=8.5cm,angle=0}\caption{a)
The numerator, $\rho $, of the X$^{-}$ Lorentzian as a function
of $V_{\text{g}}$ for six laser intensities. b) The electron
density $n_{\text{e}}$ as a function of $V_{\text{g}}$, obtained
by applying
Eq. (2) to the data of (a). Inset: $n_{\text{e}}$ as a function of $V_{\text{%
g}}$ obtained from VdP measurements.}\label{fig2}
\end{figure}

Let us address the X$^{-}$ Lorentzian lineshape, which we express as

\begin{equation}
I_{\text{X}^{-}}(E)=\frac{\rho }{(E-E_{\text{peak}}^{-})^{2}+\Gamma ^{2}}%
\text{ ,}
\end{equation}
where $E_{\text{peak}}^{-}$ and $\Gamma $ are the X$^{-}$ peak energy and
half width at half maximum, respectively. In general, a Lorentzian line is
obtained by Fourier transforming an exponentially decaying state, $\Psi (%
{\bf r})\exp (-\frac{iE_{\text{peak}}^{-}}{\text{%
h\hskip-.2em\llap{\protect\rule[1.1ex]{.325em}{.1ex}}\hskip.2em%
}}t-\frac{\Gamma }{\text{%
h\hskip-.2em\llap{\protect\rule[1.1ex]{.325em}{.1ex}}\hskip.2em%
}}t)$, and taking its absolute value squared. Thus, the numerator of the
Lorentzian $\rho =$%
h\hskip-.2em\llap{\protect\rule[1.1ex]{.325em}{.1ex}}\hskip.2em%
$^{2}|\Psi ({\bf r})|^{2}$ should depend linearly on the probability density
of the state $\Psi $, and in this particular case, on the X$^{-}$ density.
Since the X$^{-}$ density should depend linearly on the electron density $n_{%
\text{e}}$, we expect that $\rho $ should be linear in
$n_{\text{e}}$. Figure \ref{fig2}a shows $\rho $ (obtained from
the fit to the far-field spectra) as a function of the gate
voltage, $V_{\text{g}}$, for different laser
intensities. It is clearly seen that $\rho $ depends linearly on $V_{\text{g}%
}$, with a slope\ that increases linearly with laser intensity, $I_{\text{L}}
$. Thus, we can write $\rho =a[V_{\text{g}}-V_{0}(I_{\text{L}})]$, where $%
a=kI_{\text{L}}$. This linear dependence of $\rho $ on $V_{\text{g}}$ is
very significant: Using a parallel plate capacitor model for the gated 2DEG
we can express the 2DEG density as $n_{\text{e}}=C[V_{\text{g}}-V_{0}]$,
where $C$ is the geometrical capacitance between the Shottkey gate and the
2DEG layer. Hence, we conclude that that $\rho $ is indeed proportional to $%
n_{\text{e}}$, and we can obtain the 2DEG density by
\begin{equation}
n_{\text{e}}=\rho \frac{C}{a}\text{ ,}
\end{equation}
and $a$\ is simply the slope of the $\rho $ versus $V_{\text{g}}$ curve at
some arbitrary laser intensity. This is a new method to extract $n_{\text{e}}
$ in the low density regime from the PL data and the geometrical
capacitance, without having to measure $V_{0}$ and $I_{\text{L}}$.

To verify the linear dependence of $n_{\text{e}}$ on
$V_{\text{g}}$ we measured the conductance using the van der Pauw
(VdP) technique \cite{VdP}, under different illumination
intensities. The results are shown in the inset of Fig.
\ref{fig2}b. Two observations can be made: First, the density
does change linearly with gate voltage with a slope $C$, which is
the capacitance per unit area between the Shottkey gate and the
2DEG (the agreement is better than 5$\%$). The second observation
is the shift of $V_{0}$ to more positive values with increasing
laser intensity. The implication of this behavior is that the
electron density in the well is depleted under illumination by an
amount that is proportional to the laser intensity, but is
independent of the gate voltage. The probable mechanism of this
depletion process is tunneling of photo-excited electrons to the
surface, leaving the photo-excited holes trapped in the well
\cite{Glasberg}. The recombination of these holes with the 2DEG
electrons results in the reduction of the 2DEG
density. In Fig. \ref{fig2}b we apply Eq. (2) to calculate the electron density $n_{%
\text{e}}$ from the measured values of $\rho $ of Fig.
\ref{fig2}a. We get a set of parallel lines similar to those
obtained by the VdP measurements. Earlier reports have attempted
to extract the electron density from the ratio between the X and
X$^{-}$ intensities \cite{Menassen,Technion SSC}. We have
found that this method does not give an electron density that is linear in $%
V_{\text{g}}$. Furthermore, the resulting electron densities are
substantially lower than the actual ones.

\begin{figure}[h]
\centering\epsfig{file=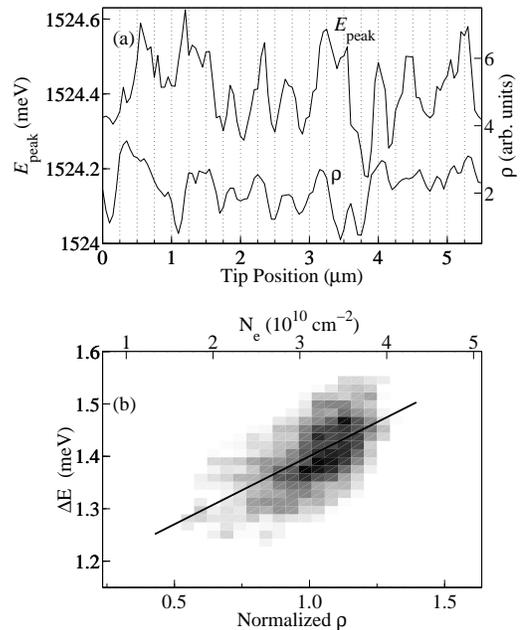,height=8.5cm,angle=0}%
\caption{a) The fluctuations in $\rho $\ and in the X peak energy
$E_{\text{peak}}$\ along a line scan of 5.5 $\mu $m at
$n_{\text{e}}=3.1\times 10^{10}$ cm$^{-2}$. Note the correlation
between the two curves. b) A two-dimensional histogram of $\rho
$\ and $E_{\text{peak}}$ for the whole scanned area ($\sim 3000$
points).} \label{fig3}
\end{figure}%
Let us try now to understand the origin of the inhomogeneous
broadening of the X line. We have shown above that the neutral
exciton peak energy, denoted as $E_{\text{peak}}$, exhibits large
spatial fluctuations (Fig. \ref{fig1}b). Examining these
fluctuations in the near-field spectra, we find that they are
strongly correlated with the fluctuations in the X$^{-}$
intensity, and hence with the local electron density. Figure
\ref{fig3}a shows the results of a line scan over $5.5$ $\mu $m
with steps of 0.05 $\mu $m. At each spatial point we extract the
values of the parameters $E_{\text{peak}}$ and $\rho $. It is
clearly seen that the fluctuations in $\rho $ are in phase with those of $E_{%
\text{peak}}$: Their minima and maxima occur at nearly the same
spatial locations. To determine the relation between the
fluctuations in the local density and in the peak energy we plot
in Fig. \ref{fig3}b a two-dimensional
histogram of the energy difference $\Delta E=E_{\text{peak}}-E_{\text{peak}%
}^{-}$ and $\rho $ for all $\sim 3000$ points in the scanned area
at one gate voltage (we normalize $\rho $ by dividing it by its
average value). Using Eq. 2 we can relate $\rho $ and
$n_{\text{e}}$, and hence express the horizontal scale in units
of electron density (upper axis). The correlation between $\Delta
E$ and $n_{\text{e}}$ is clearly visible, and emphasized by the
line, which is a linear fit for the measured $\sim 3000$ points.
It is important to emphasize that the relation between $\Delta E$
and $n_{\text{e}} $ holds locally, regions with high (low)
electron density giving rise to large (small) $\Delta E$. Since
$E_{\text{peak}}^{-}$ is nearly constant (Fig. \ref{fig1}) the
major contribution to the fluctuations in $\Delta E$ comes from
fluctuations in $E_{\text{peak}}$ \cite{Comment broadening}. We
can therefore conclude that the dominant cause of the X
inhomogeneous broadening is local density fluctuations. These
fluctuations occur on a length scale of the order of the spacer
width \cite{Efros,Nixon}, which is smaller than our spatial
resolution. Hence, each near-field point unavoidably measures a
distribution of electron densities, giving rise to inhomogeneous
broadening of the X line already in the near-field spectrum. We
note that
the broadening of the two-dimensional histogram is independent of $n_{\text{e%
}}$, and we believe that it is due to exciton diffusion,
characterized by a diffusion length of $\sim 1$ $\mu $m: Excitons
that are created in a low electron density region may diffuse to
a high density region and recombine there.%

The dependence of $\Delta E$ on $n_{\text{e}}$ was recently a subject of
intensive interest. Theoretical analysis predicts that this energy
difference should depend on the density as $\Delta E=\mu (n_{\text{e}})+E_{%
\text{B}}$, where $\mu $ is the 2DEG chemical potential and $E_{\text{B}}$
is the X$^{-}$ binding energy \cite{Hawrylak}. This dependence should be
understood as follows: When an X$^{-}$ is neutralized to form an X, the
released electron is placed at the chemical potential energy. Hence, the
cost in energy of this process is the sum of the binding energy and the
chemical potential. This relation was recently studied experimentally \cite
{HuardPRL,Yusa}. It was shown that while it holds at high densities, a
weaker dependence of $\Delta E$ on $n_{\text{e}}$ is measured in the low
density regime \cite{Yusa}. Indeed, the slope of the histogram is smaller by
more than a factor of three than that expected from theory. The fact that
the far-field inhomogeneous exciton lineshape is directly related to the
electron density fluctuations gives a powerful and simple tool for
determining these fluctuations.
\newpage
In conclusion, we have shown that the PL spectrum of a low density 2DEG can
be quantitatively understood in terms of the underlying microscopic
properties of the 2DEG. We have presented simple and novel methods to
extract the average electron density and its fluctuations from the far-field
PL spectrum.

This research was supported by the Minerva Foundation. We wish to thank
Nanonics for providing the NSOM tips.


\begin{references}
\bibitem{Kheng}  K. Kheng {\it et al.}, Phys. Rev. Lett. {\bf 71}, 1752
(1993).

\bibitem{GlebPRL}  G. Finkelstein, H. Shtrikman and I. Bar-Joseph, Phys.
Rev. Lett. {\bf 74}, 976 (1995).

\bibitem{Shields}  A. J. Shields {\it et al.}, Phys. Rev. {\bf B51}, 18049
(1995).

\bibitem{Efros}  A.L. Efros, Solid State Commun. {\bf 65}, 1281(1988) and
{\bf 70}, 253 (1989); A.L. Efros, F.G. Pikus, and V.G. Burnett, Phys. Rev.
{\bf B47}, 2233 (1993).

\bibitem{Nixon}  J.A. Nixon and J.H. Davies, Phys. Rev. {\bf B41}, 7929
(1990).

\bibitem{EytanPRL}  G. Eytan {\it et al.}, Phys. Rev. Lett. {\bf 81}, 1666
(1998).

\bibitem{NSOM Ultramicroscopy}  G. Eytan, {\it et al.}, Ultramicroscopy {\bf %
83}, 25 (2000).

\bibitem{Karrai}  K. Karrai and R. D. Grober, Appl. Phys. Lett. {\bf 66},
1842 (1995).

\bibitem{Menassen}  A. Menassen {\it et al.}, Phys. Rev. {\bf B54}, 10609
(1996).

\bibitem{VdP}  L.J. van der Pauw, Phillips Res. Rep. {\bf 13}, 1 (1958).

\bibitem{Comment VdP}  The VdP measurements were performed on a different
device, processed from the same wafer. Hence, $V_{0}$ is different.

\bibitem{Glasberg}  S. Glasberg {\it et al.}, Phys. Rev. {\bf B59}, R10425
(1999).

\bibitem{Technion SSC}  A. Ron {\it et al.}, Solid State Commun. {\bf 97},
741 (1996).

\bibitem{Comment broadening}  The width of the distribution of $\Delta E$ in
smaller than that of $E_{\text{peak}}$. This implies that some fluctuations,
which are common to X and X$^{-}$ and do not depend on electron density,
were subtracted. The most likely source of these common fluctuations is well
width fluctuations.

\bibitem{Hawrylak}  P. Hawrylak, Phys. Rev. {\bf B 44,} 3821 (1991); J. Brum
and P. Hawrylak, Comments on Cond. Matt. Phys. {\bf 18}, 135 (1997)

\bibitem{HuardPRL}  V. Huard {\it et al.}, Phys. Rev. Lett. {\bf 84}, 187
(2000).

\bibitem{Yusa}  G. Yusa, H. Shtrikman, I. Bar-Joseph, to be published in
Phys. Rev. {\bf B} (2000).
\end{references}
\end{document}